\documentclass[12pt,preprint]{aastex}
\newcommand{\kms}{km~s\ensuremath{^{-1}}}
\newcommand{\om}{\ensuremath{\Omega_m}}

\newcommand{\ola}{\ensuremath{\Omega_{\Lambda}}}
\newcommand{\kmsmpc}{{\ensuremath{{\rm km~s}^{-1}~{\rm Mpc}^{-1}}}}

\newcommand{\lcdm}{\ensuremath{\Lambda}CDM}
\newcommand{\cosmol}[3]{\ensuremath{\om = #1, \ola = #2, H_0 = #3~\kmsmpc}}
\newcommand{\lcdmparm}{\cosmol{0.27}{0.73}{71}}
\newcommand{\etal}{et~al.\/}
\newcommand{\mrc}{PKS\,1138-262}
\newcommand{\lya}{Ly\ensuremath{\alpha}}
\newcommand{\ha}{H\ensuremath{\alpha}}

\newcommand{\dms}[3]{\ensuremath{{#1}\ {#2}\ {#3}}}
\newcommand{\hms}[3]{\ensuremath{{#1}\ {#2}\ {#3}}}
\newcommand{\ie}{i.\,e.\,}
\newcommand{\eg}{e.\,g.\,}

\newcommand{\hzrgs}{HzRGs}
\newcommand{\esca}{erg~s\ensuremath{^{-1}}~cm\ensuremath{^{-2}}~\AA\ensuremath{^{-1}}}
\newcommand{\esc}{erg~s\ensuremath{^{-1}}~cm\ensuremath{^{-2}}}
\newcommand{\sqam}{arcmin\ensuremath{^{2}}}

\newcommand{\mpcc}{\,Mpc\ensuremath{^3}}

\shorttitle{Large scale structure around \mrc}
\shortauthors{Croft \etal}

\begin{document}

\title{The Filamentary Large Scale Structure around the \ensuremath{z = 2.16} Radio Galaxy \mrc}

\author{Steve Croft\altaffilmark{1}, Jaron Kurk\altaffilmark{2}, Wil van Breugel\altaffilmark{1}, S.~A.~Stanford\altaffilmark{1,3}, Wim de Vries\altaffilmark{1}, Laura Pentericci\altaffilmark{4} and Huub R\"ottgering\altaffilmark{5}}
\altaffiltext{1}{Institute of Geophysics and Planetary Physics, Lawrence Livermore National Laboratory L-413, 7000 East Avenue, Livermore, CA 94550}
\altaffiltext{2}{INAF, Osservatorio Astrofisico di Arcetri, Largo Enrico Fermi 5, 50125, Firenze, Italy}
\altaffiltext{3}{Department of Physics, University of California at Davis, 1 Shields Avenue, Davis, CA 95616}
\altaffiltext{4}{Dipartimento di Fisica, Universit\`a degli Studi Roma Tre, Italy}
\altaffiltext{5}{Leiden Observatory, Niels Bohrweg 2, NL-2333 CA Leiden, The Netherlands}

\begin{abstract}

\mrc\ is a massive radio galaxy at $z = 2.16$ surrounded by
overdensities of \lya\ emitters, \ha\ emitters, EROs and X-ray
emitters. Numerous lines of evidence exist that it is located in a
forming cluster. We report on Keck spectroscopy of candidate members of
this protocluster, including nine of the 18 X-ray sources detected by
\citet{pent:x} in this field.  Two of these X-ray sources (not counting
\mrc\ itself) were previously confirmed to be members of the
protocluster; we have discovered that an additional two (both AGN) are
members of a filamentary structure, at least $3.5$~Mpc in projection,
aligned with the radio jet axis, the 150~kpc-sized emission-line halo,
and the extended X-ray emission around the radio galaxy. Three of the
nine X-ray sources observed are lower redshift AGN, and three are
M-dwarf stars.

\end{abstract}

\keywords{galaxies: active --- galaxies: individual (\objectname{\mrc})}

\section{Introduction}

\mrc\ is a massive forming radio galaxy at $z = 2.156$ \citep{pent:98}, which is surrounded by overdensities of \lya\ emitters
\citep{pent:lyspec}, \ha\ emitters \citep{kurk:haspec}, EROs
\citep{kurk:haero} and X-ray emitters \citep{pent:x}, several of which
are spectroscopically confirmed to be close to the radio galaxy
redshift. These overdensities appear to be spatially aligned with each
other, and with the radio axis of \mrc\ (\S~\ref{sec:over}). 

At least half a dozen high redshift radio galaxies, with $2< z < 5.2$,
are now known to be located in such overdense regions \citep[protoclusters;][]{ven:02}. Candidate members of such protoclusters may be found using color selection, line emission in a narrow band targeted at the redshift of interest, or X-ray emission -- as noted above, when applied to the field of \mrc, these methods all revealed overdensities. Protoclusters have also been found using Lyman Break techniques \citep[\eg,][]{steidel:98} and deep millimeter and sub-mm observations \citep[\eg][]{ivison:00,smail:03}, and confirmed with spectroscopic follow-up. Due to the atmospheric cutoff, $z \sim 2$ is the lowest redshift at which the \lya\ selection technique will work using ground-based optical instruments, but the technique has been successfully used to detect clumps and filamentary large scale structure even at $z \sim 6$ \citep[\eg,][and references therein]{ouchi:05}.

Recent simulations suggest that the colors of massive galaxies in the local Universe can only be explained if AGN feedback quenches star formation in the host galaxy \citep{springel:05}. The link between supermassive black hole and galaxy formation \citep{hk:00} may depend critically on AGN feedback, via coupling of the mechanical power (via winds or jets) of AGN to the baryonic component of forming galaxies \citep{rawlings:03}. The mergers, galaxy harassment, and other process invoked in AGN triggering, are common in clusters, and become increasingly important as redshift increases \citep{moore:98}. High redshift clusters are dynamic regions at the nodes of the Large Scale Structure, where huge amounts of gravitational energy involved in structure formation may be dissipated. In order to understand the link between AGN, starburst and ULIRG galaxies \citep[\eg][]{nagar:03}, and the life-cycle of a typical AGN, the study of AGN in clusters and protoclusters is important. Studies of the AGN / galaxy populations, kinematics, and substructures in protoclusters provide insight into how galaxies, AGN and clusters of galaxies form and evolve. 

\mrc\ is one of the first, brightest, and lowest redshift radio galaxy protoclusters discovered
using \lya\ narrow-band selection techniques. By studying \mrc, we examine an important link between protoclusters at higher redshift, and their more evolved, virialised counterparts seen at $z \sim 1$ \citep{ford:04}, and can help constrain models of structure formation and cosmology. This, and the many X-ray emitters with unknown redshifts, motivated
us to obtain further spectra of candidate members of this protocluster using slitmasks at Keck.
The previous spectroscopically confirmed overdensities led us to suspect that even more of these targets, in addition to some of the candidate \lya\ emitters which did not yet have redshifts, would turn out to be members of the protocluster. If a large fraction of the X-ray sources turned out to be at the cluster redshift, this would suggest that the AGN activity in this cluster is unusually high, and if a large fraction of the \lya\ sources turn out to be protocluster members, this would tend to suggest that star formation is also enhanced.

\label{sec:cosmo}We assume an \lcdmparm\ cosmology \citep{wmap}. In comparing with the analyses of \citet{kurk:lyim} and \citet{pent:lyspec} note that these authors use an \cosmol{1.0}{0.0}{50} cosmology. However, due to a calculation error, the comoving volume in the cosmology of these papers should have been 2740\mpcc\ rather than the 3830\mpcc\ quoted in \citet{pent:lyspec}. In considering the comoving volume in which the objects are located, note also that, as discussed by \citet{kurk:haero}, the latest \lya-candidate catalog is based on a re-reduction of the imaging data of \citet{kurk:lyim}. The area used for detection by \citet{kurk:haero} was 38.90\,\sqam\ (not, in fact, 43.6\,\sqam\ as stated by those authors), 10\% larger than the 35.4\,\sqam\ used by \citet{kurk:lyim} and \citet{pent:lyspec}. 

In our \lcdm\ cosmology (similar to the cosmology of \citealt{kurk:haero}), the comoving area of the field used for target selection is 98\,Mpc$^2$. This yields a comoving volume of 4249\mpcc\ probed by the spectroscopic observations (in the redshift range of 2.139 -- 2.170), fortuitously rather close to the 3830\mpcc\ of \citet{pent:lyspec}. Using the redshift range of 2.110 -- 2.164 probed by the narrow-band imaging \citep{kurk:lyim} yields a comoving volume of 7405\mpcc. The conclusions of \citet{kurk:haero} are essentially unchanged, \ie, we can estimate the strength of the overdensity to be a factor $2 \pm 1$, since the comoving volume density of candidate \lya\ emitters is 3.1 times smaller than the factor six overdensity of galaxies at $z = 3.09$ discovered by \citet{steidel:00} -- see \citet{kurk:haero} for more details. The conclusions of \citet{pent:x} are also unaffected by our analysis as here the comparison to other fields is made on the basis of an areal overdensity. As noted by \citeauthor{pent:x}, the strength of the overdensity (around 50\% when compared to the Chandra Deep Fields) is similar to those found in galaxy clusters such as 3C\,295 and RX\,J0030 \citep{cappi:01}, and may, at least in part, be due to an excess of AGN in this field.

\section{Keck observations}

Our observations were made on 2004 January 19 -- 20, using the Low
Resolution Imaging Spectrometer \citep[LRIS;][]{lris} on Keck I, with
the D680 dichroic and a slit width of 1.5\arcsec. On the blue side, a
400 line / mm grism, blazed at 3400\,\AA\ was employed, giving 1.09
\AA\ / pixel and spectral resolution 8.1\,\AA. On the red side, a 400
line / mm grating, blazed at 8500\,\AA\ was used, giving 1.86 \AA\ /
pixel and spectral resolution 7.3\,\AA. This setup gives spectral coverage of $\sim 3150-9400$\,\AA, although this varies somewhat from slitlet to slitlet. Two slitmasks were observed for
9000\,s each. Seeing was $0.8$\arcsec\ during both sets of observations.

Our primary targets were randomly selected from those X-ray emitters of
\citet{pent:x} and candidate \lya\ emitters of \citet{kurk:haero} which did not yet have spectroscopically confirmed
redshifts. \citet{kurk:haero} select LEG (\lya-emitting galaxy) candidates from a 38.90\,\sqam\ field on the basis of excess narrow versus broad band flux, from which they calculate a continuum subtracted narrow band flux, $F = F_{NB} - F_{B}$ (where $F_{NB}$ is the flux measured in a narrow band filter with central wavelength 3814\,\AA\ and FWHM 65\,\AA, and $F_{B}$ is the flux measured in Bessel $B$-band). They also calculate a corresponding rest frame equivalent width, $EW_0$ (assuming that the narrow band excess is due to \lya\ emission at $z = 2.16$). Their values of $F$ and $EW_0$ for our targets are shown in Table~\ref{tab:ids}.

Our
slitmasks were designed by attempting to maximise the number of primary
targets per mask. Gaps were then filled with secondary targets: five
objects confirmed as members of the protocluster by
\citet{pent:lyspec}. In total we observed 31 targets
(Table~\ref{tab:ids}). The slitmasks were designed with slitlets ranging in length from $\sim 15 - 60$\arcsec, in all cases adequate for good sky subtraction. 

The data were reduced in the standard manner using {\sc
bogus}\footnote{\url{http://astron.berkeley.edu/$\sim$dan/homepage/bogus.html}}
in {\sc iraf}, and spectra extracted in a 1.3\arcsec-wide aperture.

\section{Properties of the spectroscopic targets}

Table~\ref{tab:ids} summarises the results for all of our spectroscopic
targets. Four targets were undetected and three showed continuum but no
clear emission lines. Five showed an obvious single emission line which
we were unable to unambiguously identify; since continuum is seen
blueward of the line in these cases it is likely to be [\ion{O}{2}]
3727. Assuming this to be the case, we assign redshifts in
Table~\ref{tab:ids} but mark these instances with a question mark.
In total we obtained secure redshifts for nineteen new targets which we classify 
on the basis of the broadness of spectral lines; [\ion{O}{3}] / H$\beta$,
[\ion{O}{2}] / [\ion{O}{3}] and other relevant line ratios (where these
could be measured) and / or X-ray characteristics, 
as discussed below. Spectral lines were fit with Gaussian profiles using {\sc splot} in {\sc iraf}. For L968 (X3), where broad and narrow components are present, deblending was performed by fitting with multiple Gaussians. Measured parameters for the emission lines in those objects believed to be members of the \mrc\ protocluster are listed in Table~\ref{tab:lines}. The observed-frame equivalent width of the fitted profile, $EW_{\lambda}$, was used to compute the rest-frame equivalent width, $EW_0 = EW_{\lambda} \times (\lambda_{rest} / \lambda_{obs})$, which is shown in the table. In some cases our measured equivalent widths differ from the estimates of \citet{kurk:haero}; this is because it is difficult to get an accurate measure of the line flux and the faint continuum from filters which are relatively broad compared to the spectral features of interest. All lines discussed below are in emission unless otherwise noted.

\begin{figure}
\centering
\includegraphics[width=\linewidth]{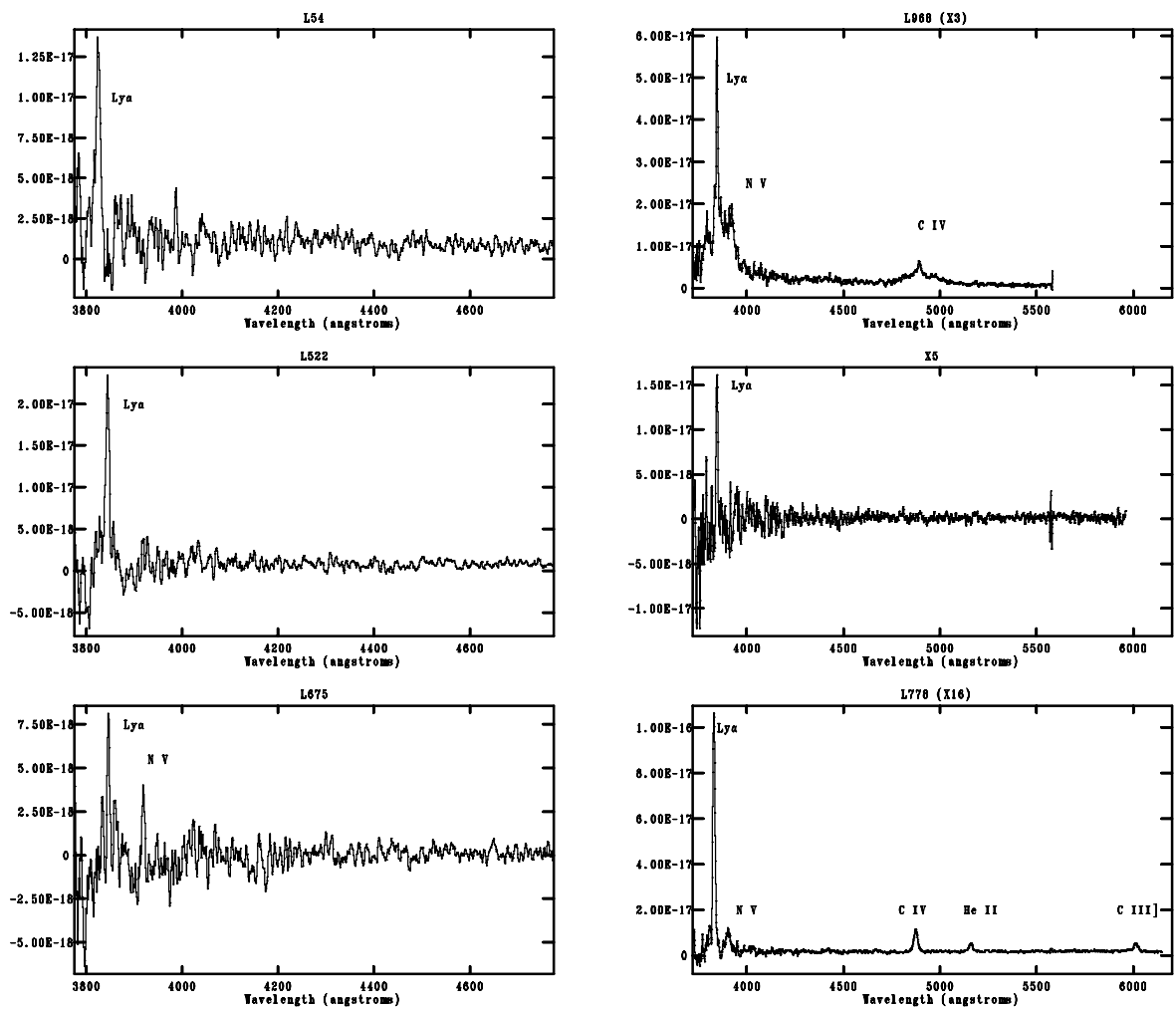}
\caption{\label{fig:lyanew}The LRIS-B (blue side) emission spectra of the objects from Table~\ref{tab:ids} confirmed to be in the $z \sim 2.16$ cluster, smoothed with a boxcar of width five pixels. Four of these objects were already spectroscopically confirmed by \citet{pent:lyspec}; the two newly-confirmed $z \sim 2.16$ \lya-emitters are X5 and L778 (X16). The X-ray emitters are shown plotted over a wider wavelength range, as unlike the LEG candidates, some show emission lines over this wider region. The y-axes show flux density in \esca. }
\end{figure}

\paragraph{L54} Previously confirmed at $z = 2.143$ by \citet{pent:lyspec}. We obtain $z = 2.145$. Our spectrum is shown in Fig.~\ref{fig:lyanew}.

\begin{figure}
\centering
\includegraphics[width=0.75\linewidth]{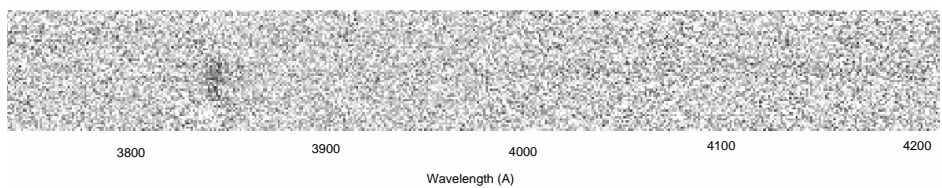}
\caption{\label{fig:l522}Two-dimensional spectrum of L522; the \lya\ emission (observed wavelength 3844 \AA) is spatially extended over a wider area than the continuum, showing the presence of the \lya\ halo around this object.}
\end{figure}

\paragraph{L522} Previously confirmed at $z = 2.166$ by \citeauthor{pent:lyspec}\ We obtain $z = 2.161$ (see Fig.~\ref{fig:lyanew}). The \lya\ emission is extended  (Fig.~\ref{fig:l522}) over 2.7\arcsec\ (23~kpc) across the slit (PA = 82\degr), due to the \lya\ halo which encompasses this object and the radio galaxy \citep{kurk:halo}. Indeed, in the narrow-band image, the halo is seen to be several hundred kiloparsecs in extent -- this is not uncommon for such objects \citep[\eg,][]{vanojik:96}, and is proposed to be due to the infall of primordial gas and galaxy ``building blocks'' such as Lyman Break Galaxies \citep{reuland:03}. 

\paragraph{L675} Previously confirmed at $z = 2.163$ by \citeauthor{pent:lyspec} (we measure $z = 2.162$; see Fig.~\ref{fig:lyanew}), and also detect \ion{N}{5} 1240.

\paragraph{L778 (X16)} A new spectroscopic confirmation at $z = 2.149$, showing broad \lya\ (rest-frame deconvolved FWHM 890\,\kms), \ion{N}{5} 1240, \ion{C}{4} 1549, \ion{He}{2} 1640, \ion{C}{3}] 1909, [\ion{C}{2}] 2326 and \ion{Mg}{2} 2798 (Fig.~\ref{fig:lyanew}).

\begin{figure}
\centering
\includegraphics[bb=37 325 575 430,clip=true,width=0.75\linewidth]{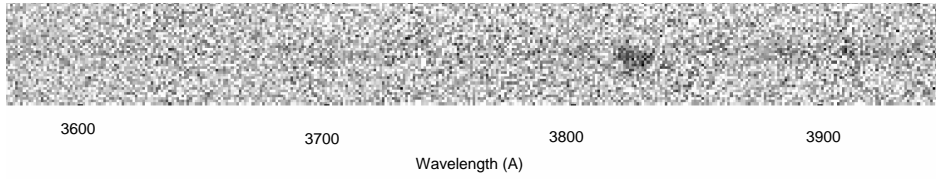}
\caption{\label{fig:l891}Two-dimensional spectrum of L891. The emission line at observed wavelength 3828 \AA\ is most probably \lya, but the continuum bluewards of this line casts doubt on this interpretation. The center of the line is offset by around 0.7\arcsec\ from the continuum, suggesting that they may in fact be due to two separate objects.}
\end{figure}

\paragraph{L891} Listed as a $z = 2.147$ \lya\ emitter by \citeauthor{pent:lyspec}, this object was observed in {\em both} of our masks, and shows continuum blueward of the supposed \lya\ line at 3828\,\AA, as well as an emission line at 7176\,\AA. We were unable to identify a plausible combination of emission lines arising from a single object at any redshift which could give rise to this line combination. We note, however, that in the 2D spectra (Fig.~\ref{fig:l891}), the ``\lya'' appears offset slightly (around 0.7\arcsec) from the continuum emission. This may be a LEG at the cluster redshift (we measure $z = 2.148$), while the continuum and 7176\,\AA\ emission line are continuum and [\ion{O}{2}] 3727 from a foreground object at $z = 0.925$ (which also shows some evidence of a continuum break in around the right place for Ca H + K).\label{sec:l891}

\paragraph{L968 (X3)} Previously confirmed at 2.183 (with caution that \lya\ appears self-absorbed). We obtain $z = 2.162$, and see only slight self-absorption blueward of \lya\ (Fig.~\ref{fig:lyanew}). This AGN shows broad and narrow \lya, broad and narrow \ion{C}{4} 1549, and broad \ion{N}{5}. The \ion{C}{4} emission was fit with two Gaussian components, and the broad and narrow \lya\ and \ion{N}{5} were fit with three Gaussian components.

\paragraph{X2, X11 \& X17} H$\alpha$ through H$\zeta$ at $z = 0$ are seen, along with molecular absorption bands. Comparison with the spectroscopic standards of \citet{mdwarf} shows these three objects to be M-dwarf stars.

\paragraph{X5} Faint but significant \lya\ emission (rest-frame deconvolved FWHM 400\,\kms); a new spectroscopic confirmation at $z = 2.162$ (Fig.~\ref{fig:lyanew}). This object is also listed as an ERO (ERO\,226; $I - K = 5.0$) by \citet{kurk:haero}. Although it is difficult to say, based on only the observed \lya\ line, whether this object is a starburst or AGN, the soft X-ray to optical flux ratio \citep{pent:x}, in addition to an inferred soft X-ray luminosity ($4 \times 10^{43}$ erg s$^{-1}$)  much too high for a starburst, provide strong evidence for the presence of an AGN. 

\paragraph{X18} This $z = 0.436$ Seyfert 2 shows strong [\ion{O}{2}] 3727, [\ion{O}{3}] 4959, 5007, [\ion{Ne}{3}] 3869, [\ion{Ne}{5}] 3426 and H$\alpha$ lines. It is the nucleus of a galaxy undergoing a merger as seen on the VLT image. 

\section{Discussion}
 
All of the nine X-ray selected targets were spectroscopically detected
and classified. Six of these were AGN, three of which are members of
the protocluster $z \sim 2.2$ (including one reconfirmation).  Two of
these were also LEG candidates, showing again the efficacy of the narrow-band imaging 
selection method.  Of the three lower-redshift AGN, one is at $z =
1.117$ (a member of the $z = 1.16$ spike discussed below).

The three remaining X-ray targets are M-dwarf stars. This is an
unusually large fraction.  Among the 42 objects with spectroscopic
classifications in the Chandra Deep Field South catalog of \citet{cdfs} only one X-ray
source brighter than $10^{-15}$ \esc\ is an M-dwarf. So the X-ray
overdensity in the \mrc\ field is at least in part due to an excess of
M-dwarfs in the foreground. However, with five out of 18 X-ray sources
(including \mrc) now confirmed to be AGN this still suggests a higher
AGN fraction in the \mrc\ protocluster than in local clusters
\citep[see][]{pent:x} and in the fields of other \hzrgs\ \citep{overzier:05}.

Of the 24 targets selected on the basis of their \lya\ excess, five were
at the redshift of the
protocluster (including four previously confirmed), seven were undetected or showed only faint continuum, and the remainder are most likely
foreground galaxies. Four of these show a single emission line in the range
7927 -- 8235\,\AA, which, if interpreted as [\ion{O}{2}]~3727, places them at
$z = 1.16 \pm 0.05$. The remaining five are at $z < 1$.
The apparent overdensity at $z = 1.16$, which also includes L479 and
X4 in addition to the four single-emission-line objects, 
makes it all the more necessary to obtain high quality spectra 
of more targets in this field to determine whether they belong to the system at $z = 1.2$, the 
\mrc\ protocluster at $z = 2.2$, or are unassociated with either.
For example, the starburst galaxy L479 at $z = 1.171$ lies at the
center of a 20\arcsec-long string of six EROs \citep{kurk:haero}, suggesting that at least
some of the EROs may be physically associated with the $z = 1.16$
overdensity. Apparently, at this redshift, starburst galaxies 
with strong [\ion{O}{2}] emission are also picked up
by the \lya\ selection process (at quite low significance) due to
bluening of their continuum by young stars. 

The pure LEG-selected targets in our multislit observations did not
reveal additional \mrc\ protocluster members (with the exception of
L778, which is also an X-ray emitter). This is unlike the
high success rate (up to 70\%)
of spectroscopically confirmed $z \sim 2.2$ LEGs reported by \citet{pent:lyspec}. This is probably because objects with 
large $EW$ (estimated from narrow-band photometry) are easiest to detect spectroscopically. 
However, these also tend to be
the faintest objects in B-band (by definition) and large $EW$ {\em
combined with low expected \lya\ flux} creates a signal-to-noise
problem. Indeed, we see that our four non-detections are those listed
by \citet{kurk:haero} as having large $EW$, and comparatively low
\lya\ flux (Table~\ref{tab:ids}).

To dispel any concern that our observations were less sensitive than
those of \citet{pent:lyspec}, note that the five of their confirmed $z
\sim 2.2$ LEGs which we observed are reconfirmed by our observations
(with the possible exception of L891, as noted above --- which does in
any case show a line which could plausibly be \lya). In all cases
except L968 (see discussion above), values of $1 + z$ obtained are in
agreement at the $\sim 0.2\%$ level. It seems therefore that our
success rate in confirming \lya\ emitters is lower simply due to our
targets being those which are intrinsically fainter and / or closer to
the cutoff of the selection criteria. We note that the \lya\ selection
technique has shown great success in other fields, particularly at
higher redshift, and it is perhaps not surprising that our success rate
is somewhat lower here due to redshifted \lya\ being towards blue
optical wavelengths. Here spectroscopic efficiency begins to drop, and
the measurement of broad-band fluxes below the \lya\ line or Lyman limit as an additional 
selection criterion is impossible due to the atmospheric cutoff.

\section{Conclusions}

There is still a good deal of evidence that \mrc\ is located in a
forming cluster; this conclusion is bolstered by our confirmation of
two new objects at the same redshift as the radio galaxy. That one of
these is also an ERO lends credence to the hypothesis that at least
some of the ERO overdensity may be attributed to the protocluster.
However, as noted by \citet{steidel:00}, the narrow-band (Equivalent
Width) selection technique is effective but also difficult to
quantify.  Sensitive, wide spectral wavelength coverage is essential,
the more so because there is a non-trivial chance that overdense fields
at high redshifts maybe confused by other overdense fields in the
foreground \citep{projeffect}; another example of such a superposition of unrelated overdensities is seen by \citet{francis:04}. Curiously, \citeauthor{francis:04} also detect an object showing two emission lines which they are unable to indentify, similar to L891 as discussed above.

\begin{figure}
\epsscale{0.7}
\plotone{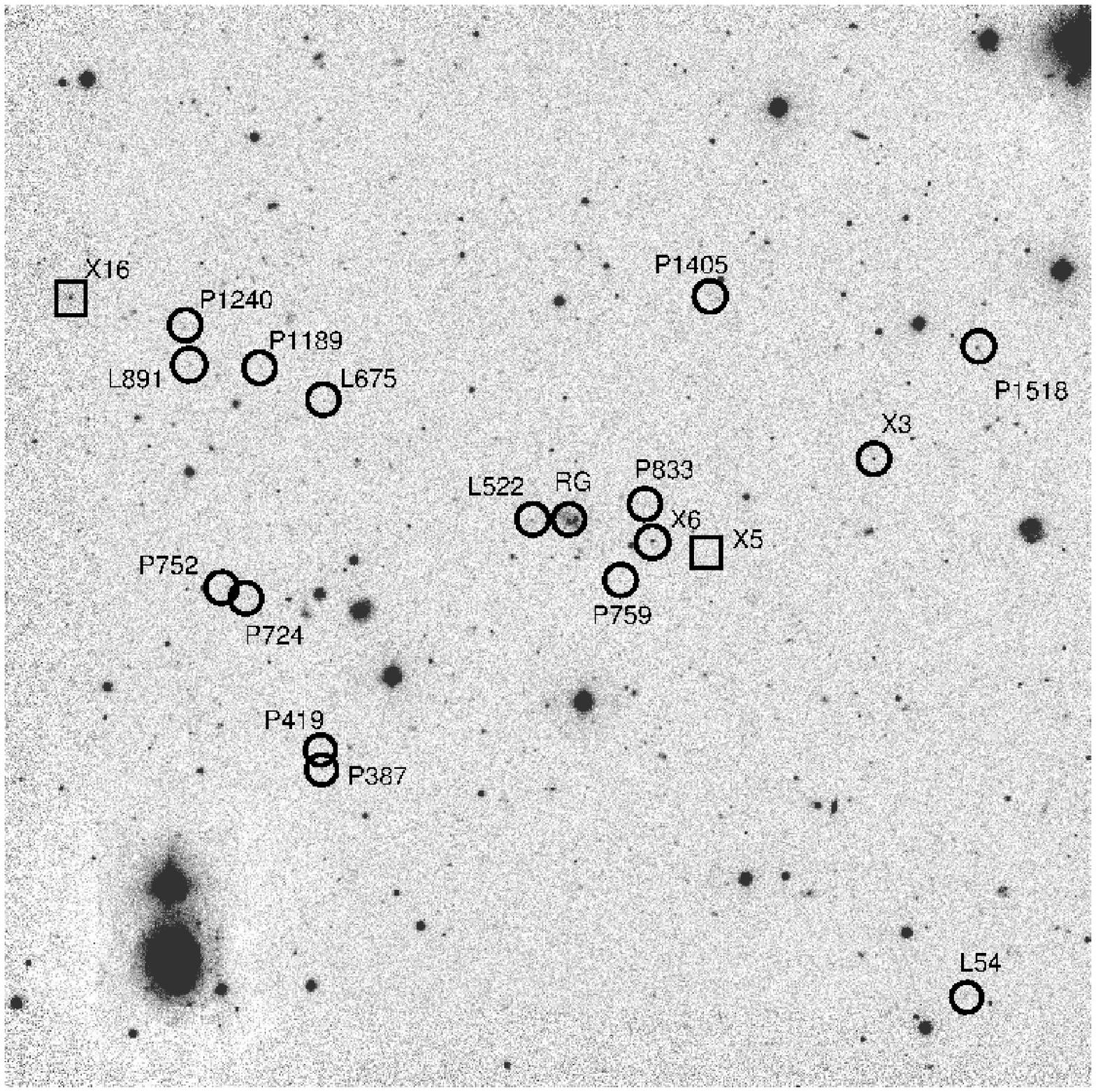}
\caption{\label{fig:clus}A $6.8\arcmin \times 6.8\arcmin$ narrow-band (3814\,\AA) image of the field from VLT/FORS \citep{kurk:lyim}. Previously confirmed protocluster members are indicated by circles (including L891, but see text). The `L' and `X' objects are labelled as in Table~\ref{tab:ids}. Additionally the radio galaxy \mrc\ is labelled `RG', and the `P' objects are labelled as in \citet{pent:lyspec}. Note that the four `L' objects, plus X3 (L968) are reobservations of targets from \citet{pent:lyspec}; the two new confirmations from this paper are X5 and X16, and are indicated by squares.} 
\end{figure}

The confirmation that around one third of the X-ray sources in this field are associated with the protocluster ties in well with the conclusions of \citet{pent:x} that there is a 50\% excess of X-ray sources in the field of \mrc\ compared to the Chandra Deep Fields. The high AGN fraction in \mrc\ suggests that the AGN were probably triggered at around the same time, presumably by the ongoing formation of the 
protocluster. This supports models where AGN feedback is an important component of the early phases of galaxy and cluster formation. 
\label{sec:over} 
 Intriguingly, our two new spectroscopic confirmations
lie along the line of an apparent overdensity of
spectroscopically-confirmed \lya\ emitters (Fig.~\ref{fig:clus}), in
the same direction as the overdensity of X-ray emitters, radio axis of
the central galaxy, extended X-ray emission around \mrc, and the
general distribution of \ha\ emitters in the cluster \citep{pent:x}.
This is not entirely surprising if what we are seeing is a filament of
the Large Scale Structure associated with this forming galaxy cluster.

\acknowledgments

The data presented herein were obtained at the W.\ M.\ Keck Observatory, which is operated as a scientific partnership among the California Institute of Technology, the University of California and the National Aeronautics and Space Administration. The Observatory was made possible by the generous financial support of the W.\ M.\ Keck Foundation. Work was performed under the auspices of the U.\ S.\ Department of Energy, National Nuclear Security Administration by the University of California, Lawrence Livermore National Laboratory under contract No.\ W-7405-Eng-48.

% here beginneth the figures / tables

%\begin{figure}
%\centering
%\includegraphics[bb=85 325 530 475,clip=true]{f4.eps}
%\caption{\label{fig:lyanew}The emission spectra of the two newly-confirmed $z \sim 2.16$ \lya-emitters. The y-axis shows flux density in \esca. X5 (right panel) has been smoothed with a boxcar of width five pixels. }
%\end{figure} 

\clearpage

\begin{deluxetable}{lllllll}
\rotate
\tablewidth{0pt}
\tabletypesize{\scriptsize}
%\tabletypesize{\tiny}
\tablecaption{\label{tab:ids}Results of spectroscopy for the LEG candidate (`L') and X-ray (`X') samples.}
\tablehead{
\colhead{ID \tablenotemark{a}} & \colhead{RA} & \colhead{Dec} & \colhead{$z$} & \colhead{$EW_0$ \tablenotemark{b}} & \colhead{$F$ \tablenotemark{c}} & \colhead{Notes}
}
\startdata
L46 & \hms{11}{40}{33.89} & \dms{-26}{32}{13.4} & \nodata & 31.6 & 3.4 & continuum\\
L54 \tablenotemark{d, e} & \hms{11}{40}{37.15} & \dms{-26}{32}{08.3} &  2.145 & 24.2 & 7.1 & LEG P7\\
L73 & \hms{11}{40}{37.75} & \dms{-26}{31}{55.5} & 0.671 & 34.5 & 3.1 & [\ion{O}{3}] 4959, 5007; H$\beta$; starburst?\\
L127 & \hms{11}{40}{48.21} & \dms{-26}{31}{32.4} & \nodata & 18.9 & 2.2 & continuum\\
L184 & \hms{11}{40}{36.63} & \dms{-26}{31}{04.1} & \nodata & 22.5 & 4.4 & continuum\\
L286 & \hms{11}{40}{47.92} & \dms{-26}{30}{31.5} & \nodata & 265.0 & 2.6 & ND\\
L361 & \hms{11}{40}{36.88} & \dms{-26}{30}{08.8} & 0.913 & 23.6 & 4.3 & [\ion{O}{2}] 3727; [\ion{O}{3}] 5007; broad H$\gamma$; H$\beta$ coincides with skyline; Seyfert 1?\\
L365 & \hms{11}{40}{58.08} & \dms{-26}{30}{09.4} & 1.127? & 21.8 & 3.7 &7927\,\AA\ emission line\\
L366 & \hms{11}{40}{51.65} & \dms{-26}{30}{08.0} & 0.861 & 29.5 & 5.0 & H$\beta$; \ion{C}{3}] 1909; strong [\ion{O}{2}] 3727; AGN?\\
L470 & \hms{11}{40}{59.71} & \dms{-26}{29}{35.0} & 1.210? & 49.9 & 3.9 & 8235\,\AA\ emission line\\
L479 & \hms{11}{40}{51.05} & \dms{-26}{29}{31.1} & 1.171 & 25.0 & 3.3 & [\ion{O}{2}] 3727; weak H$\delta$; starburst?\\
L484 & \hms{11}{40}{45.52} & \dms{-26}{29}{30.0} & \nodata & 96.6 & 1.4 & ND\\
L522 \tablenotemark{d, e} & \hms{11}{40}{49.40} & \dms{-26}{29}{09.6} & 2.161 & 36.0 & 20.0 & LEG P856\\
L565 & \hms{11}{40}{56.56} & \dms{-26}{26}{13.6} & 0.496? & 28.2 & 3.9 & 5574\,\AA\ emission line\\
L674 & \hms{11}{40}{45.95} & \dms{-26}{28}{35.7} & \nodata & 269.4 & 2.6 & ND\\
L675 \tablenotemark{d, e} & \hms{11}{40}{55.29} & \dms{-26}{28}{24.3} & 2.162 & 117.8 & 3.1 & LEG P1612\\
L739 & \hms{11}{40}{57.42} & \dms{-26}{27}{07.8} & 1.125? & 20.6 & 6.3 & 7919\,\AA\ emission line\\
L877 & \hms{11}{40}{54.04} & \dms{-26}{28}{01.1} & 0.863 & 31.6 & 4.3 & [\ion{O}{2}] 3727; narrow H$\beta$; weak H + K abs; starburst?\\
L891 \tablenotemark{e} & \hms{11}{40}{59.07} & \dms{-26}{28}{10.5} & 2.146? & 25.4 & 4.7 & P1557; see text\\
L941 & \hms{11}{40}{44.04} & \dms{-26}{28}{33.5} & 0.800 & 32.0 & 3.3 & H$\beta$; [\ion{O}{3}] 4959, 5007; starburst?\\
L942 & \hms{11}{40}{49.84} & \dms{-26}{28}{29.0} & \nodata & 55.3 & 2.9 & ND\\
L980 & \hms{11}{40}{42.31} & \dms{-26}{28}{51.3} & 1.171? & 113.1 & 4.7 & 8093\,\AA\ emission line\\
   X2 & \hms{11}{40}{38.83} & \dms{-26}{29}{10.3} & 0.000 & \nodata & \nodata & M-dwarf\\
   X3 $\equiv$ L968 \tablenotemark{d, e} & \hms{11}{40}{39.69} & \dms{-26}{28}{44.9} & 2.162 & 26.6 & 9.8 & AGN P1687\\
   X4 & \hms{11}{40}{44.21} & \dms{-26}{31}{29.8} & 1.117 & \nodata & \nodata & [\ion{Ne}{4}] 2424; [\ion{O}{2}] 3727; [\ion{Ne}{3}] 3869; AGN\\
   X5 \tablenotemark{d} & \hms{11}{40}{44.46} & \dms{-26}{29}{20.6} & 2.162 & \nodata & \nodata & ERO, AGN\\
   X9 & \hms{11}{40}{52.83} & \dms{-26}{29}{11.2} & 1.512 & \nodata & \nodata & \ion{C}{4} 1549; \ion{He}{2} 1640; \ion{C}{3}] 1909; \ion{Mg}{2} 2798; BL AGN\\
   X11 & \hms{11}{40}{54.66} & \dms{-26}{29}{28.1} & 0.000 & \nodata & \nodata & M-dwarf\\
   X16 $\equiv$ L778 \tablenotemark{d} & \hms{11}{41}{02.41} & \dms{-26}{27}{45.1} & 2.149 & 314.2 & 67.6 & AGN\\
   X17 & \hms{11}{41}{02.99} & \dms{-26}{27}{34.1} & 0.000 & \nodata & \nodata & M-dwarf\\
   X18 \tablenotemark{f} & \hms{11}{41}{03.93} & \dms{-26}{30}{48.4} & 0.436 & \nodata & \nodata & Seyfert 2\\
%\cutinhead{Serendipitous detections}
%s470.2 & \nodata & \nodata & \nodata & \nodata & 8227 \\
%s73.2 & \nodata & \nodata & 0.671? & \nodata & Hb, OIII\\
%s891.2 & \nodata & \nodata & \nodata & \nodata & 8306\\
%s980.2 & \nodata & \nodata & \nodata & \nodata & 7263\\
%s54.2 & \nodata & \nodata & 0 & \nodata & star\\
%sx2.2 & \nodata & \nodata & \nodata & \nodata & 8103\\
\enddata
\tablenotetext{a}{Catalog number from \citet{kurk:haero}.}
\tablenotetext{b}{Rest-frame equivalent width (in \AA) estimated from the continuum-subtracted narrow-band observations of \citet{kurk:haero}, assuming that excess narrow-band emission is due entirely to \lya\ emission at $z = 2.16$.}
\tablenotetext{c}{Continuum-subtracted narrow-band flux (in 10$^{-17}$ \esc) measured by \citet{kurk:haero}.}
\tablenotetext{d}{Confirmed cluster member. See Table~\ref{tab:lines} for properties of the observed emission lines.}
\tablenotetext{e}{\ Listed as spectroscopically confirmed protocluster member by \citet{pent:lyspec}. Note that we find that one of these objects (L891) may not in fact be at the cluster redshift. The catalog numbers in \citet{pent:lyspec} do not match those in \citet{kurk:haero}; where applicable the \citeauthor{pent:lyspec} IDs (`P') are given in the Notes column.}
\tablenotetext{f}{Hard X-ray source. All other X-ray sources in the table are soft \citep{pent:x}.}
\tablecomments{Objects which were undetected in spectroscopic observations are marked ``ND''. All listed lines were detected in emission. Right ascension and declination are given in hours, minutes, and seconds, and degrees, arcminutes, and arcseconds, respectively, for equinox J2000.}
\end{deluxetable}

% X ids are Pentericci numbers

\begin{deluxetable}{lllllll}
%\rotate
\tablewidth{0pt}
\tabletypesize{\scriptsize}
%\tabletypesize{\tiny}
\tablecaption{\label{tab:lines}Properties of the observed emission lines for protocluster members.}
\tablehead{
\colhead{ID} & \colhead{Line} & \colhead{$\lambda_{rest}$ (\AA)} & \colhead{$\lambda_{obs}$ (\AA)} & \colhead{FWHM (\kms)} & \colhead{Flux (10$^{-17}$ \esc)} & \colhead{$EW_0$ (\AA)}
}
\startdata
L54 & \lya\ & 1216 & 3824 & $730 \pm 20$ & $15.7 \pm 0.2$ & $57.3 \pm 0.6$ \\
L522 & \lya\ & 1216 & 3844 & $230 \pm 20$ & $20.8 \pm 0.1$ & $57.0 \pm 0.6$ \\
L675 & \lya\ & 1216 & 3845 & $200 \pm 30$ & $6.4 \pm 0.1$ & \nodata \tablenotemark{a} \\
 & \ion{N}{5} & 1240 & 3919 & $200 \pm 30$ & $4.1 \pm 0.1$ & \nodata \tablenotemark{a} \\
L891 \tablenotemark{b} & \lya\ $z=2.148$? & 1216? & 3828 & $770 \pm 20$ & $23.9 \pm 0.2$ & $26.2 \pm 0.2$ \tablenotemark{c}\\
& [\ion{O}{2}] $z=0.925$? & 3727? & 7176 & $220 \pm 30$ & $6.2 \pm 0.2$ & $49 \pm 1$ \tablenotemark{c}\\
X3 $\equiv$ L968 & \lya\ broad & 1216 & 3845 & $4790 \pm 40$ & $36.9 \pm 0.2$ & $157 \pm 0.9$ \\
 & \lya\ narrow & 1216 & 3845 & $122 \pm 20$ & $118.4 \pm 0.8$ & $49.0 \pm 0.3$ \\
 & \ion{N}{5} & 1240 & 3923 & $4041 \pm 40$ & $74.0 \pm 0.6$ & $89.8 \pm 0.6$ \\
 & \ion{C}{4} broad & 1549 & 4895 & $12180 \pm 280$ & $48.2 \pm 0.8$ & $63 \pm 2$ \\
 & \ion{C}{4} narrow & 1549 & 4884 & $1070 \pm 50$ & $5.7 \pm 0.3$ & $6.1 \pm 0.3$ \\
X5 & \lya\ & 1216 & 3845 & $ 400 \pm 30 $ & $ 18.0 \pm 0.1 $ & \nodata \tablenotemark{a} \\
X16 $\equiv$ L778 & \lya\ & 1216 & 3829 & $890 \pm 20$ & $158.4 \pm 0.2$ & $294.4 \pm 0.4$ \\
 & \ion{N}{5} & 1240 & 3901  & $2030 \pm 20$ & $24.0 \pm 0.2$ & $36.7 \pm 0.3$ \\
 & \ion{C}{4} & 1549 & 4873  & $1430 \pm 20$ & $24.8 \pm 0.2$ & $54.5 \pm 0.3$ \\
 & \ion{He}{2}& 1640 & 5157  & $1360 \pm 30$ & $9.3 \pm 0.2$ & $19.3 \pm 0.4$ \\
 & \ion{C}{3}] & 1909 & 6012  & $1220 \pm 40$ & $10.0 \pm 0.2$ & $19.6 \pm 0.4$ \\
 & [\ion{C}{2}] & 2326 & 7334 & $1730 \pm 210$ & $4.0 \pm 0.5$ & $13 \pm 2$ \\
 & \ion{Mg}{2} & 2798 & 8822 & $960 \pm 70$ & $6.9 \pm 0.4$ & $11.5 \pm 0.6$\\
\enddata
\tablenotetext{a}{No continuum detected.}
\tablenotetext{b}{\ Line identifications uncertain; this object may not be at the cluster redshift (see text).}
\tablenotetext{c}{Assuming $z = 2.148$.}
\tablecomments{In contrast to Table~\ref{tab:ids}, the values in this table are those measured from our Keck spectra.}
\end{deluxetable}


\begin{thebibliography}{}
\bibitem[Best(2000)]{best:00} Best, P.~N.\ 2000, \mnras, 317, 720 
\bibitem[Cappi et al.(2001)]{cappi:01} Cappi, M., et al.\ 2001, \apj, 548, 624
\bibitem[Carilli et al.(1998)]{carilli:98} Carilli, C.~L., Harris, D.~E., Pentericci, L., Rottergering, H.~J.~A., Miley, G.~K., \& Bremer, M.~N.\ 1998, \apjl, 494, L143 
\bibitem[Chambers, Miley, \& van Breugel(1987)]{chamb:87} Chambers, K.~C., Miley, G.~K., \& van Breugel, W.\ 1987, \nat, 329, 604 
\bibitem[Ford et al.(2004)]{ford:04}Ford, H., et al. 2004, in Penetrating Bars through Masks of Cosmic Dust: The Hubble Tuning Fork Strikes a New Note, ed. D. L. Block, et al. (New York: Springer); astro-ph/0408165
\bibitem[Francis et al.(2004)]{francis:04} Francis, P.~J., 
Palunas, P., Teplitz, H.~I., Williger, G.~M., \& Woodgate, B.~E.\ 2004, 
\apj, 614, 75
\bibitem[Haehnelt \& Kauffmann(2000)]{hk:00} Haehnelt, M.~G., \& Kauffmann, G.\ 2000, \mnras, 318, L35
\bibitem[Hill \& Lilly(1991)]{hill:91} Hill, G.~J.~\& Lilly, S.~J.\ 1991, \apj, 367, 1 
\bibitem[Ivison et al.(2000)]{ivison:00} Ivison, R.~J., Dunlop, J.~S., Smail, I., Dey, A., Liu, M.~C., \& Graham, J.~R.\ 2000, \apj, 542, 27 
\bibitem[Kirkpatrick, Henry, \& McCarthy(1991)]{mdwarf} Kirkpatrick, J.~D., Henry, T.~J., \& McCarthy, D.~W.\ 1991, \apjs, 77, 417 
\bibitem[Kurk et al.(2000)]{kurk:lyim} Kurk, J.~D., et al.\ 2000, \aap, 358, L1 
\bibitem[Kurk et al.(2004a)]{kurk:haero} Kurk, J.~D., Pentericci, 
L., R{\" o}ttgering, H.~J.~A., \& Miley, G.~K.\ 2004, \aap, 428, 793
 \bibitem[Kurk et al.(2004b)]{kurk:haspec} Kurk, J.~D., Pentericci, 
L., Overzier, R.~A., R{\" o}ttgering, H.~J.~A., \& Miley, G.~K.\ 2004, 
\aap, 428, 817 
\bibitem[Kurk(2003)]{kurk:halo}Kurk, J.~D.\ 2003, PhD Thesis, Leiden Observatory
\bibitem[Lacey \& Cole(1993)]{lacey:93} Lacey, C.~\& Cole, S.\ 1993, \mnras, 262, 627 
\bibitem[Moore et al.(1998)]{moore:98} Moore, B., Lake, G., \& Katz, N.\ 1998, \apj, 495, 139
 \bibitem[Nagar et al.(2003)]{nagar:03} Nagar, N.~M., Wilson, A.~S., Falcke, H., Veilleux, S., \& Maiolino, R.\ 2003, \aap, 409, 115 
\bibitem[Oke et al.(1995)]{lris} Oke, J.~B., et al.\ 1995, \pasp, 107, 375
\bibitem[Ouchi et al.(2005)]{ouchi:05} Ouchi, M., et al.\ 2005, \apjl, 620, L1 
\bibitem[Overzier et al.(2005)]{overzier:05} Overzier, R.~A., Harris, D.~E., Carilli, C.~L., Pentericci, L., R{\" o}ttgering, H.~J.~A., 
\& Miley, G.~K.\ 2005, \aap, 433, 87
\bibitem[Pentericci et al.(1998)]{pent:98} Pentericci, L., R{\"o}ttgering, H.~J.~A., Miley, G.~K., Spinrad, H., McCarthy, P.~J., van 
Breugel, W.~J.~M., \& Macchetto, F.\ 1998, \apj, 504, 139 
\bibitem[Pentericci et al.(2000)]{pent:lyspec} Pentericci, L., et al.\ 2000, \aap, 361, L25 
\bibitem[Pentericci et al.(2002)]{pent:x} Pentericci, L., Kurk, J.~D., Carilli, C.~L., Harris, D.~E., Miley, G.~K., \& R{\"o}ttgering, H.~J.~A.\ 2002, \aap, 396, 109
\bibitem[Rawlings(2003)]{rawlings:03} Rawlings, S.\ 2003, New Astronomy Review, 47, 397 
\bibitem[Reuland et al.(2003)]{reuland:03} Reuland, M., et al.\ 2003, \apj, 592, 755
\bibitem[Simpson \& Rawlings(2002)]{simpson:02} Simpson, C.~\& Rawlings, S.\ 2002, \mnras, 334, 511
 \bibitem[Smail et al.(2003)]{smail:03} Smail, I., Scharf, C.~A., Ivison, R.~J., Stevens, J.~A., Bower, R.~G., \& Dunlop, J.~S.\ 2003, \apj, 599, 86 
\bibitem[Spergel et al.(2003)]{wmap} Spergel, D.~N., et al.\ 2003, \apjs, 148, 175 
\bibitem[Springel et al.(2005)]{springel:05} Springel, V., Di Matteo, T., \& Hernquist, L.\ 2005, \apjl, 620, L79 
\bibitem[Steidel et al.(1998)]{steidel:98} Steidel, C.~C., Adelberger, K.~L., Dickinson, M., Giavalisco, M., Pettini, M., \& Kellogg, M.\ 1998, \apj, 492, 428
\bibitem[Steidel et al.(2000)]{steidel:00} Steidel, C.~C., Adelberger, K.~L., Shapley, A.~E., Pettini, M., Dickinson, M., \& Giavalisco, M.\ 2000, \apj, 532, 170
\bibitem[Szokoly et al.(2004)]{cdfs} Szokoly, G.~P., et al.\ 
2004, \apjs, 155, 271 
\bibitem[Venemans et al.(2002)]{ven:02} Venemans, B.~P., et al.\ 2002, \apjl, 569, L11 
\bibitem[van Haarlem, Frenk, \& White(1997)]{projeffect} van Haarlem, M.~P., Frenk, C.~S., \& White, S.~D.~M.\ 1997, \mnras, 287, 817 
\bibitem[van Ojik et al.(1996)]{vanojik:96} van Ojik, R., Roettgering, H.~J.~A., Carilli, C.~L., Miley, G.~K., Bremer, M.~N., \& Macchetto, F.\ 1996, \aap, 313, 25 
\end{thebibliography}
\end{document}